\documentstyle[aps,epsfig]{revtex}

\def\be{\begin{equation}}
\def\ee{\end{equation}}
\def\beq{\begin{eqnarray}}
\def\eeq{\end{eqnarray}}
\def\brt{\rightarrow}
\def\rt{\rightarrow}

\begin{document}

\title {Phase Diagram of a Two-Species Lattice Model with a Linear Instability}
\author{Sriram Ramaswamy$^{1,}$ \cite{byjnc,email1}, 
Mustansir Barma$^{2,}$\cite{byjnc,email2},
Dibyendu Das$^{2,}$\cite{email3} and Abhik Basu$^{1,3,}$\cite{email4}}
\address{$^1$Centre for Condensed Matter Theory, Department of Physics, Indian
Institute of Science, Bangalore 560012, India,\\
$^2$Department of Theoretical Physics, Tata Institute of Fundamental 
Research, Homi Bhabha Road, Mumbai 400005, India,\\
$^3$Poornaprajna Institute of Scientific Research, Bangalore, India}
\maketitle

\begin{abstract}

{We discuss the properties of a one-dimensional lattice model of a driven 
system with two species of particles in which the mobility of one
species depends on the density of the other. This model was
introduced by Lahiri and Ramaswamy in the context of sedimenting
colloidal crystals, and its continuum version was shown to exhibit an
instability arising from linear gradient couplings.
In this paper we review recent progress in understanding the full phase
diagram of the model. There are three phases. In the first, the
steady state can be determined exactly along a representative locus using
the condition of detailed balance.
The system shows phase separation of an exceptionally robust sort,
termed strong phase separation, which survives at all temperature.
The second phase arises in the threshold case where
the first species evolves independently of the second, but the
fluctuations of the first influence the evolution of the second,
as in the passive scalar problem. The second species then shows
phase separation of a delicate sort, in which long range order
coexists with fluctuations which do not damp down in the large-size
limit. This fluctuation-dominated phase ordering is associated with
power law decays in cluster size distributions and a breakdown of the
Porod law. The third phase is one with a uniform
overall density,  and along a representative locus the steady state 
is shown to have product measure form.  Density fluctuations are
transported by two kinematic waves, each involving
both species and coupled at
the nonlinear level. Their dissipation properties are governed by
the symmetries of these couplings, which depend on
the overall densities.  In the most interesting case,
the dissipation of the two modes is characterized by different
critical exponents, despite the nonlinear coupling.}

\end{abstract}


\section{Introduction}

Much recent progress in the general area of nonequilibrium steady states has
come about by focusing on simple driven lattice-gas models which have the
same symmetries as the more complex physical situations they represent.  For
instance, the asymmetric exclusion process --- a simple model of hard-core
particles performing biased random walks --- is representative of a class of
current-carrying states and growth processes involving a single scalar
field \cite{zia4}.
Now, there are several examples of physical processes which involve two
coupled scalar fields \cite{ertas14,ertas24,kardar4,bara14,bara24,lahiri14};
for such situations, the corresponding lattice model
involves two sets of Ising variables, with the time evolution of each
species depending on the instantaneous configuration of the other. Such 
systems can display interesting, new types of steady states, and in
this paper we discuss their occurrence within a recently introduced
lattice model of coupled fields.

This model was initially proposed  by Lahiri and Ramaswamy \cite{lahiri14}
(hence referred to
as the LR model) in connection with collective effects in
sedimentation, the process by which heavier particles
settle in a lighter fluid.
Interactions between sedimenting particles are mediated by the fluid
\cite{crowley14,crowley24,ramaswamy4}, and
give rise to a coupled dynamics of two fields, namely the concentration (the
particle number density) and the tilt field (the orientation of the
principal axes of the particle distribution). A distinctive aspect of the
inter-field coupling in this case is that within
a hydrodynamic description, the coupling involves {\it linear terms},
in addition
to more customary nonlinear coupling terms. 
This essential aspect is captured by the 1-d LR lattice model.
Linear couplings can also arise in other contexts, for instance in  
reduced models of magnetohydrodynamics \cite{abjkbsr}, 
drifting flux lattices in superconductors \cite{aditi}, and the motion of
stuck and flowing grains on a sandpile \cite{anita}.  The essential aspect 
of linear coupling is captured by the 1-d LR lattice model. In the 
sedimentation problem,
analysis of the linearized hydrodynamic equations points to an instability
in a certain regime, but leaves open the question of the ultimate steady
state reached by the system.  The lattice model allows this question
to be answered;
the instability can be studied by varying the transition rates within the 
model,
and the  resulting steady state can be analysed and characterized
\cite{lahiri14,lahiri24}.

The instability in question is towards large-scale clustering, and the 
question
arises to what extent the phenomenon resembles phase separation, familiar
from equilibrium statistical mechanics.  In fact, the nature
of phase separation in nonequilibrium steady states is an area of great
interest at present \cite{mukamel14,evansreview,hinrichsen}, and  the
study of the LR model proves quite illuminating in this
context. The unstable phase turns out to be strongly phase separated,
much as in the three-component model of Evans et al 
\cite{mukamel24,mukamel34}. The phase separation
is robust, and survives at all temperatures. The behaviour is quite different,
however, at the threshold of the instability, which corresponds to
the particularly simple situation  in which one of the two fields evolves
autonomously while the other is driven by the fluctuations of the first,
as in the passive scalar problem. There is phase ordering in this case as
well, but it is quite
delicate. Fluctuations of the order parameter are very strong, but do not
destroy the ordered phase \cite{das4}; rather, they change its character in an
essential way, leading to the breakdown of familiar characteristics such as
the Porod law for the decay of scaled correlations. Finally, moving
away from the threshold, even the disordered
regime of the model has some surprises to offer. In this regime, the density
is homogeneous on a macroscopic scale, but fluctuations are transported
ballistically by two sets of kinematic waves \cite{lighthill4}. 
The waves are coupled nonlinearly,
yet under special symmetry-determined conditions (i.e., on a special 
locus in parameter space) the dissipation of
each  is governed by a distinct critical exponent \cite{fourauth}. 
This is the first such
instance to come to light in a system with nonvanishing nonlinear coupling.
In the kinematic wave phase in general, the equations are a one-dimensional 
reduced model for the dynamics of small fluctuations of a 
flux lattice drifting through a type II superconductor \cite{aditi}, and 
also serve as a generalised Burgers equation for magnetohydrodynamics (MHD) 
\cite{abjkbsr}.   

In the remainder of this section, we discuss the occurrence of the
instability within a linearized hydrodynamic description, and then
define the LR lattice model and briefly discuss its phase diagram. In the
sections that follow, we discuss respectively strong phase separation
(in the unstable regime), fluctuation-dominated phase separation (at the
threshold), and kinematic waves and their decay (in the stable regime).

\subsection{Hydrodynamic Description and Instabilities}

The hydrodynamic equations of a coupled-field system involve
the density fluctuations $\delta
\rho_{1,2}(x,t) \equiv \rho_{1,2}(x,t) - \rho_{1,2}^o$ of the two species
in question. When dealing with conserved fields, the equations governing
their time evolutions may be derived from the continuity equations 
$\partial  \rho_1/\partial t + \partial J_1/\partial x = 0$ and
$\partial  \rho_2/\partial t + \partial J_2/\partial x = 0$, where
$J_1(\delta \rho_1, \delta \rho_2)$ and $J_2(\delta \rho_1, \delta \rho_2)$
are the currents of particles of species 1 and 2 respectively.
We may proceed by writing each of the currents $J_1$ and
$J_2$ in terms of a systematic part $J_{1,2}^{sys}$, a diffusive part $-D
\partial \rho_{1,2}(x,t)/\partial x$, and noise $\eta_{1,2}$. The coupling
between the two species arises from the systematic parts, for each of
$J_{1}^{sys}$ and $J_{2}^{sys}$ depend both on $\delta \rho_1$ and $\delta
\rho_2$.  We obtain the requisite equations on expanding in powers of the
density fluctuations.  We find it convenient to write these equations
in terms of the integrated density fields $h_{1,2}(x,t)
=\int^x \delta \rho(x',t) dx'$.  Then, to second order, we obtain
\beq
{\partial h_1 \over \partial t} &=& c_{11} {\partial h_1\over \partial
x} + c_{12} {\partial h_2 \over \partial x} + D_1 {\partial^2 h_1
\over \partial x^2} + 
\lambda_1 \left({\partial h_1\over \partial
x}\right)^2 + \mu_1 \left({\partial h_2 \over \partial x}\right)^2 +
\nu_1 \left({\partial h_1\over \partial x}\right) \left({\partial h_2
\over \partial x}\right) + \eta_1 (x,t) \nonumber \\ [2mm]
{\partial h_2 \over \partial t} &=& c_{21} {\partial h_1\over \partial
x} + c_{22} {\partial h_2 \over \partial x} + D_2 {\partial^2 h_2
\over \partial x^2} + 
\lambda_2 \left({\partial h_1\over \partial
x}\right)^2 + \mu_2 \left({\partial h_2 \over \partial x}\right)^2 +
\nu_2 \left({\partial h_1\over \partial x}\right) \left({\partial h_2
\over \partial x}\right) + \eta_2 (x,t). \nonumber \\ [2mm]
\label{coupled}
\eeq

Let us first examine the effect of keeping only the linear first derivative
terms on the right hand side of the above equations. These terms, which are 
of
primary interest to us here, were absent on symmetry grounds in 
coupled-field
problems involving
directed polymer motion and surface growth, considered respectively in
references \cite{ertas14,ertas24,kardar4} and \cite{bara14,bara24}.
Let the eigenvectors of the $2\times 2$ matrix $\pmatrix{c_{11} & c_{12} \cr
c_{21} & c_{22}}$ be $e_+$ and $e_-$, and let the corresponding
eigenvalues be $c_+$ and $c_-$. There are three cases to consider:

\begin{enumerate}
\item[{(A)}] {\it Stable:}
If $c_+$ and $c_-$ are real (i.e. if $\Delta \equiv (c_{11} -
c_{22})^2 + 4 c_{12} c_{21} > 0$), they represent  the speeds of
two waves. The two waves involve the eigenvectors $e_+$ and $e_-$,
each of which is
composed of a linear combination of $h_1$ and $h_2$.
To proceed, rewrite
Eqs. \ref{coupled} in terms of $e_+ (x,t)$ and $e_-(x,t)$. Notice
that in the rest frame of each wave, the other wave moves with a
finite speed $|c_+ - c_-|$; it is impossible to make both stationary by
a Galilean shift.  The nonlinear terms describe the couplings between the
two waves which govern their dissipation.

\item[{(B)}] {\it Threshold:}  The discriminant
$\Delta$ vanishes if one of the off-diagonal terms, say $c_{12}$,
is zero. At the linear level, this would imply that the field $h_1$ evolves
autonomously, but (since $c_{21} \ne 0$) the time evolution of the
`passive scalar' $h_2$ is influenced by $h_1$. There is a tendency for
particles of type 2 to be driven together.  In the corresponding
lattice model, this system shows phase separation accompanied,
however, by large-scale fluctuations.

\item[{(C)}] {\it Unstable:}
There is an instability
if $c_+$ and $c_-$ pick up an imaginary part (i.e. if $\Delta < 0$).
The solutions of the linear equations then have an unbounded exponential
growth of fluctuations. This instability signals the advent of a new
state which is qualitatively different from the statistically
homogeneous state assumed at the outset.  By studying the corresponding
lattice model in the unstable regime, we find that the system
undergoes macroscopic phase separation of a robust sort.

\end{enumerate}

\subsection {Lattice Model and Phase Diagram}

The Lahiri Ramaswamy model \cite{lahiri14} is the lattice counterpart 
of the continuum
equations discussed above, and analysing its phases helps us understand
the behaviour of the system in each of the cases (A), (B) and (C) above.

The LR model is defined in terms of two sets of variables $\{\sigma_i\}$
and $\{\tau_{i - {1 \over 2}}\}$ which reside on two interpenetrating 
sublattices;
the former occupy the integer sites and the latter the half-integer bond
locations of a one-dimensional lattice with $L$ sites. 
Each $\sigma_i$ and $\tau_{i-{1 \over 2}}$ is an
Ising variable taking on values $\pm 1$.
They represent discrete versions of the density and tilt fields in
the sedimentation problem: If $\sigma_i$ is
$1$, there is a particle ($+$) at site $i$, and if $\sigma_i = -1$, there is
no particle ($-$). The variable
$\tau_{i - {1 \over 2}} = 1$ and $-1$,
implies two values $/$ and $\backslash$ of the local tilt respectively. A typical
configuration of the full system is thus:
$+\backslash -/-/+\backslash -/+/+/+\backslash -$.

Both sets of variables are conserved, i.e.   $\sum \sigma_i$ and
$\sum \tau_{i-{1 \over 2}}$  and the associated densities
$\rho_1^o=\sum (1+\sigma_i)/2L$ and $\rho_2^o=\sum (1+\tau_{i-{1 \over 2}})/2L$ 
are constant. The linear terms in the hydrodynamic equations
lead us to consider a $\tau$-dependent local field which guides the
$\sigma$-current and {\it vice versa}. Thus, for instance,
the Kawasaki exchange dynamics of the adjacent spins $\sigma_i$ and
$\sigma_{i+1}$ occurs at a rate which depends on $\tau_{i+{1 \over 2}}$.
The moves and the corresponding rates are depicted below:

\begin{eqnarray}
\left[1\right]~~& + \backslash - ~~~\brt ~~~ - \backslash 
+&~~~~~~~~r_1
\nonumber \\
\left[2\right]~~& - \backslash +~~~ \brt~~~ + \backslash 
-&~~~~~~~~r_2
\nonumber \\
\left[3\right]~~&- / + ~~~\brt~~~ + / -&~~~~~~~~r_1 \nonumber \\
\left[4\right]~~&+ / - ~~~\brt~~~ - / +&~~~~~~~~r_2 \nonumber \\
\left[5\right]~~&/ - \backslash ~~~~\brt~~~ \backslash - 
/&~~~~~~~~p_2
\nonumber \\
\left[6\right]~~&\backslash - / ~~~\brt~~~ / - 
\backslash&~~~~~~~~p_1
\nonumber \\
\left[7\right]~~&\backslash + / ~~~\brt~~~ / + 
\backslash&~~~~~~~~p_2
\nonumber \\
\left[8\right]~~&/ + \backslash ~~~\brt~~~ \backslash + 
/&~~~~~~~~p_1
\label{moves}
\end{eqnarray}

The rates may be written succinctly as
\begin{eqnarray}
W(\sigma_i \leftrightarrow \sigma_{i+1}; \tau_{i+{1 \over 2}}) =
{(r_1 + r_2) \over 2} - {(r_1 - r_2) \over 4}{\tau_{i+{1 \over 2}}}
(\sigma_i - \sigma_{i+1}) \nonumber \\
W(\tau_{i-{1 \over 2}} \leftrightarrow \tau_{i+{1 \over 2}}; \sigma_i) =
{(p_1+p_2) \over 2} + {(p_1 - p_2) \over 4}{\sigma_i}(\tau_{i-{1 \over 2}} -
\tau_{i + {1\over 2}}).
\label{rates}
\end{eqnarray}
In a simple correspondence to a hydrodynamic model, discussed in 
section IV below, the coefficients in the continuum description (Eq. \ref{coupled})
depend on the rates $p_1, p_2, r_1, r_2$, as also on the overall densities 
$\rho_1^o$ and $\rho_2^o$
of the two species.  This correspondence allows us to demarcate each
of the regimes (A), (B) and (C) discussed above in terms of the parameters
of the lattice model.

Throughout the rest of the paper, we will assume that $r_1>r_2$.
Thus particles tend to move preferentially downhill, and holes
uphill. Holding the ratio $r_1/r_2$ fixed, it is then interesting to ask for
the effect of varying $p_1/p_2$. The result is summarized in the phase
diagram of Fig. 1, for the half-filled case $\sum \sigma_i = \sum \tau_i 
=0$.  We discuss the characteristics of each phase in brief.

\underline{$p_1 > p_2$} (the light grey region in Fig. 1):
In the corresponding continuum equation, the diagonal coefficients
$c_{11}$ and $c_{22}$ vanish, while $c_{12}$ and $c_{21}$ both have the same
sign.
Thus $\Delta$ is negative, corresponding to the unstable case (C).
Since $p_1>p_2$, a local peak ($/\backslash$) tends to become a
valley ($\backslash /$) if a particle resides on it,
while a valley tends to convert into a hill if a hole sits on it. These
moves act in concert with the  motion of particles
down slopes (determined by $r_1>r_2$), to promote
segregation of both kinds of spins.  In fact, using the condition of
detailed balance, the steady state can be determined
exactly along the the line
$r_1/r_2 = p_1/p_2$ shown in Fig. 1 \cite{lahiri24}. The state
shows phase segregation
of a  particularly strong sort, as discussed in section 2.

\underline{$p_1 = p_2$} (the thick central line in Fig. 1): This is the threshold case,
with $c_{12}=0$ leading to vanishing $\Delta$.
It describes a semi-autonomous
problem in which the $\tau$'s evolve independently of
the $\sigma$'s; they undergo a symmetric exclusion process. The
$\sigma$'s on the other hand obey the dynamics of Eq. \ref{rates}.
The system shows a particularly delicate sort of phase separation,
which is accompanied by large-scale fluctuations in the steady state
\cite{das4,das34}.

\begin{figure}[tb]
\begin{center}
\leavevmode
\epsfig{figure=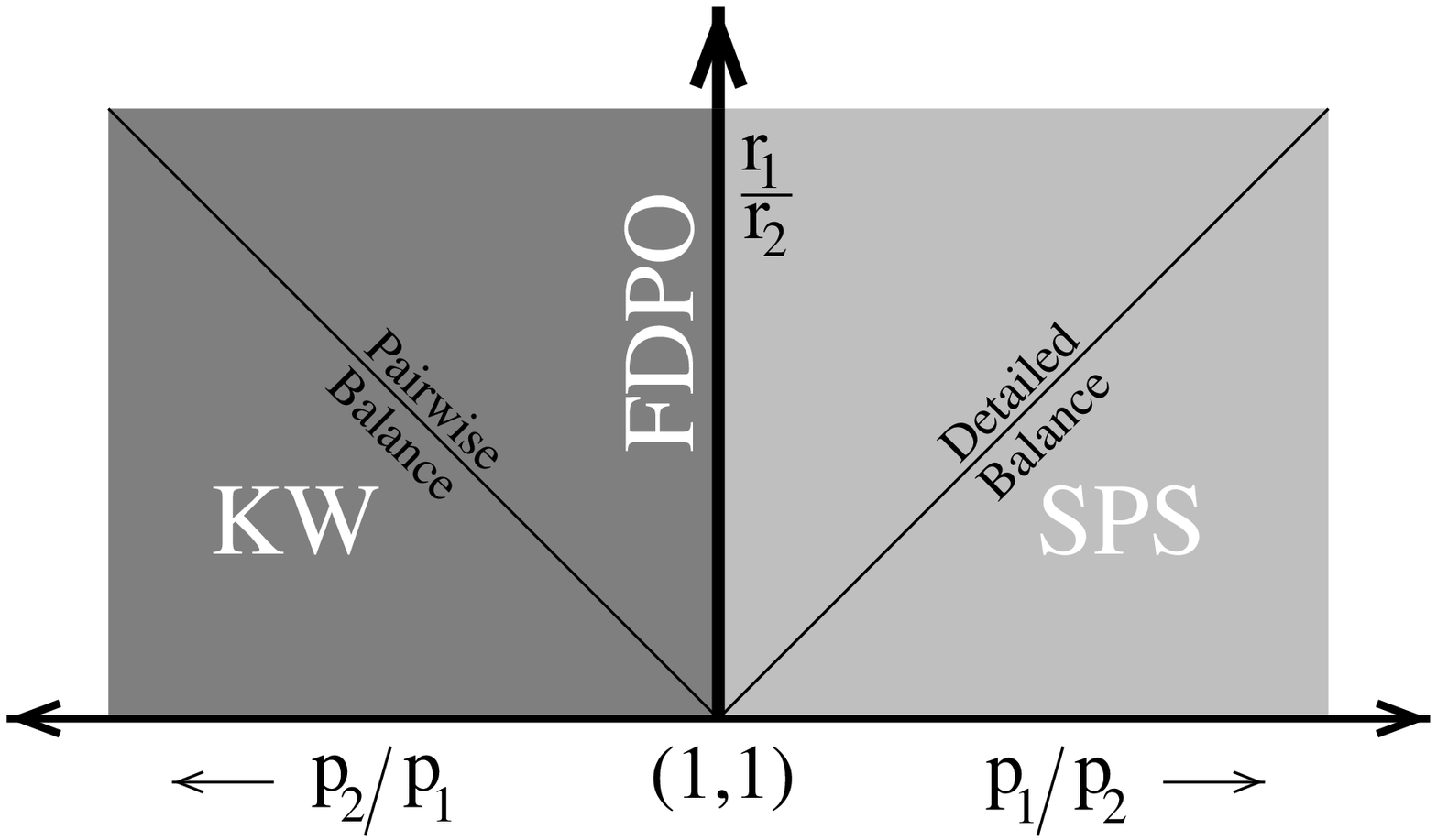,width=11cm}
\end{center}
\label{fig:phases}
\end{figure}  	 

\noindent {{\bf Figure 1}:The phase diagram of the LR model 
at half-filling $(\sum \sigma_i
= \sum \tau_{i-1/2} = 0)$.  The strongly phase separated phase (SPS)
is separated from the kinematic wave phase (KW) by the threshold,
semi-autonomous line $p_1=p_2$ where there is fluctuation-dominated phase
ordering (FDPO). The steady state can be found exactly along the line
$r_1/r_2=p_1/p_2$ (using the condition of detailed balance) and along
$r_1/r_2=p_2/p_1$ (using the condition of pairwise balance).}
\vskip 0.4cm

\underline{$p_1 < p_2$} (the dark grey region in Fig. 1): In this case
$c_{12}$ and $c_{21}$ have opposite signs, so that $\Delta$ is positive.
The tendency of  particles to slide
downhill is countered by the tendency of particle-rich valleys to become 
hills.
The resulting steady state has a uniform macroscopic density and slope, 
while
density fluctuations are carried by two kinematic waves, each corresponding 
to
a mode which involves both $\sigma$'s and $\tau$'s.
The steady state can be found exactly
along the line $r_1/r_2=p_2/p_1$ shown in Fig. 1, by using the condition
of pairwise balance \cite{barma4}.
Interestingly, the symmetries of the mode evolution equations can be tuned
by varying the overall densities, and it is thereby possible to
achieve a situation in which the  dissipation of the two modes
have different dynamical exponents in spite of being coupled at
the nonlinear level \cite{fourauth}.

\section{ Strong Phase Separation}

In the unstable phase of the LR model ($p_1 > p_2; ~~r_1>r_2$, the region
marked SPS in Fig. 1), there
is a bootstrap effect at work: the tendency of $\sigma$ particles to
fall downhill
into valleys works in tandem with the tendency of particle-rich regions
of the $\tau$ landscape
to become valleys themselves. This results in the formation of ever
deeper valleys and large-scale clustering of particles in them,
leading ultimately to a phase separated state.

A similar sort of phase separation has been shown to occur in a three species
ABC model in which a cyclic rule determines preferential rates of
exchanges of adjacent particles \cite{mukamel24,mukamel34}.
For the case with an equal number of
A, B and C particles, the steady state was found exactly, using the 
condition
of detailed balance.

In the LR model too, the exact steady state can be found \cite{lahiri24} 
through a detailed balance argument in the
case $\sum \sigma_i=
\sum \tau_i = 0$, $r_1/r_2=p_1/p_2$ (Fig. 1). This happens because
although the problem is defined in terms of transition rates,
it is possible to construct a Hamiltonian with respect to which
detailed balance is valid. This Hamiltonian has the form
\be
{\cal H} = \epsilon \sum^N_{k=1} h_k \{\tau\} \sigma_k
\label{hamilt}
\ee
where $h_k\{\tau\}$ is  a height field (the discrete analog of $h(x)$ in
Eq. \ref{coupled}) defined by
\be
h_k \{\tau\} = \sum^k_{j=1} \tau_{j-1/2} .
\label{height}
\ee
The steady state is thus described by an equilibrium Boltzmann
measure, and a configuration $\{\sigma,\tau\}$ has weight
$\exp(-{\cal H}\{\sigma,\tau\}/T)$ where $T$ is the temperature.
This can be seen by noting that the energy change on interchanging
neighbouring $\sigma$ spins is
\be
\Delta E_\sigma \equiv \Delta E (\sigma_i \leftrightarrow
\sigma_{i+1}) = \epsilon \tau_{i+{1\over 2}} (\sigma_i - \sigma_{i+1})
,
\label{exchange}
\ee
while the change of energy under exchange of $\tau$ spins is given by
a similar expression with $\sigma$'s and $\tau$'s interchanged.
Thus the ratio of Boltzmann weights of the configurations after and
before an interchange of spins,
say  $\sigma_i \leftrightarrow \sigma_{i+1}$,  is $exp(-2\Delta
E_{\sigma}/T)$.  This equals the ratio of forward to backward
transition rates if the ratio $\epsilon/T$ is related to the rates $p_1$
and $p_2$ by \be {p_2 \over p_1} = e^{-2 \epsilon/T}
\label{balance}
\ee
The ratio $r_2/r_1$ of forward and backward transition rates involving the
interchange of adjacent $\tau$'s is also related to the same
weight factor. Thus, provided $r_1/r_2=p_1/p_2$ holds, the steady state
weight of a configuration is proportional to $exp(-{\cal H})/T$.
Notice that ${\cal H}$ involves long-ranged couplings between spins
$\sigma_i$ and $\tau_{j-1/2}$.

The ground state exhibits complete phase
separation, corresponding to all $\sigma = 1$ particles being at the
bottom of the potential well formed by phase-separated $\tau$ spins. 
Pictorially,
the state is
\[
- \backslash - \backslash - \backslash - \backslash + \backslash +
\backslash + \backslash + / + / + / + / - / - / - /
\]
There are two interfaces across which there is a change of $\sigma$ spin 
values,
and two
more involving the $\tau$'s. The spacing between the four interfaces
is $L/4$, and each is completely sharp in the limit $T \rt 0$.
The effect of nonzero temperature is to smear out each
of the four interfaces over a finite length. The interfacial profile can
be calculated \cite{mukamel34}, but for our purpose here it suffices to
note that the interface width is of the order of $T$, and has basically no
effect on the bulk of the spins.
Thus in an infinite sample, phase separation survives at all $T$.
This unusual behaviour can be traced to the fact that
the energy grows super-extensively ($\sim$ (length)$^2$), and thus dominates
over the entropy at all temperatures. This allows
the strictures against phase separation in equilibrium 
one-dimensional systems \cite{landau} to be avoided.

When the filling is changed away from $\sum_i\sigma_i=
\sum_i\tau_i=0$, the condition of detailed balance no longer holds,
and there is no description in terms of an effective
Hamiltonian. Nevertheless, kinetic arguments can be given to show that
phase separation survives even when the filling is changed. The
extreme robustness of the phenomenon leads us to call it Strong Phase
Separation.

The approach to the steady state is exceptionally slow.
It occurs through a coarsening process which involves thermal activation
over large barriers, so we expect a coarsening length growing 
logarithmically in time. Logarithmically slow coarsening was found in
the ABC model as well \cite{mukamel24}.

In concluding this section, we mention a related model which has quite
different behaviour. Recall that in the LR model, the mobility of
each spin species depends on the local density of the other species.
What if the mobility depends not on density, but a higher derivative of the
density?  A case in which the coupling is to the second derivative of
the density can be analysed through detailed balance
considerations, but this time, the Hamiltonian has short-ranged interactions
\cite{turkey}. There is thus no phase ordering in this model at nonzero
$T$, in contrast to the strong phase separation exhibited by the LR model
in the unstable regime.

\section{ Fluctuation-dominated Phase Ordering}

In the marginal case $p_1=p_2, ~~~ r_1>r_2$ (the line FDPO in Fig. 1),
the evolution of the $\tau$'s
proceeds independently of the $\sigma$'s. The fluctuating $\tau$ landscape
provides a source of nonequilibrium noise on the $\sigma$'s which tend
to fall downwards along local slopes and cluster in the valleys.
Since the strong landscape fluctuations never cease, the clustering
is never as complete as in the case of
strong phase separation. Rather, as we will see below, the
character of long-range order is strongly modified by fluctuations,
hence the appellation Fluctuation-dominated Phase Ordering.

Starting from an initial random arrangement of the $\sigma$ particles,
the extent of clustering can be estimated quite simply \cite{das4}: In any
fluctuating surface characterized by a dynamical exponent
$z$, there are rearrangements of the profile over
length scales $\sim t^{1/z}$ in time $t$. This sets the scale
for the base lengths of new valleys which form in time $t$.
In the case at hand, where the scaling properties of the surface are those
of the Edwards-Wilkinson model \cite{EW4}, we have $z=2$. This valley base 
length
should set the scale for spatial clustering of particles, so we expect
the equal time correlation function to follow the scaling form
\beq
\langle \sigma_{o}(t) \sigma_{o+r}(t) \rangle
= C(r/{\cal L}(t))~~~~~~~~~~{\rm with}~~{\cal L} \sim t^{1/z} .
\label{splt}
\eeq
This is confirmed by numerical studies. This sort of
scaling behaviour is a general
characteristic of phase ordering dynamics, which describes the evolution from an
initially disordered state to one which is phase separated \cite{bray}.
However, the scaling function here is quite different from that in usual 
cases.
Normally, with a conserved order parameter, $ C(y)$ drops linearly
for small $y \equiv r/{\cal L}(t)$, i.e. $C(y) \approx C_o (1-c_1 |y|)$ 
as $y \rt 0$.
By contrast, we find $ C$ shows a cusp for small values of $|y|$, i.e.
\beq
C(y) \approx C_o (1-c_1 |y| ^{\alpha})
\label{nonporod}
\eeq
with $\alpha \simeq 0.5$ \cite{das4}.
This is significant, since a linear dependence on $|y|$ (which goes under the
name of the Porod law) is a simple consequence of having phases separated
by sharp interfaces \cite{bray}. 
The breakdown of this law is the first pointer to the
unusual character of the state that is developing.

In steady state, $ C$ again shows a similar scaling form, with the
coarsening scale ${\cal L}(t)$ being replaced by the system size $L$. The 
cusp
for small argument remains unaltered. Turning to one-point correlation
functions, a suitable measure \cite{schmitt} for our conserved spin system
is the magnitude of the Fourier components of the density profile
\begin{equation}
Q(k) = |{{1 \over L} {\sum_{j=1}^L} e^{ikj} n_j}|,~~~~~~~~~~
k = {2{\pi}m \over L}.
\label{qk}
\end{equation}
where $ n_j=(1+\sigma_j)/2$ and $m$ runs over $1,...,L-1$. A
signature of an ordered state is that
in the thermodynamic limit, the average values $\langle Q(k) \rangle$ 
approach
zero for all nonzero $k$, but $Q(k \rt 0)$ remains finite. We numerically
monitored these averages  $\langle \cdots
\rangle$ over the ensemble of steady state configurations
and found that $\langle Q(k) \rangle \rt 0$
in the thermodynamic limit for fixed $k \ne 0$, but
$\langle Q^* \rangle$ approaches a nonzero value, where
$Q^*= Q(k={2\pi \over L})$.

$\langle Q^* \rangle$ provides a quantitative measure
of phase separation.  We find that $\langle Q^* \rangle \simeq 
0.18$
indicating that the steady state is ordered, though not perfectly (a 
perfectly
ordered state has $Q^* \simeq 0.32$). As a function of time, $Q^*$ fluctuates
strongly, reflected in the fact that the distribution $P(Q^*)$ remains 
broad,
with a  root-mean-squared deviation that remains finite 
even as $L \rt \infty$.
One may ask whether phase ordering disappears when, in the course
of its fluctuations,  the value of $Q^*$ falls
close to zero.  The answer is in the negative.  We observe that a dip
in $Q^*$ is accompanied by a simultaneous rise in the value of either
$Q(m=2)$ or $Q(m=3)$. This implies that whenever the system loses a
single large cluster (making $Q^*$ small) either two or three such
clusters appear in its place (making the values of $Q(m)$ for some small
$m$ go up).  Thus the system remains far from the disordered state,
and always has a few large particle clusters which are of macroscopic
size $\sim L$.  A numerical study shows that the average size of the 
largest particle cluster $\simeq 0.14 L$.

We observe further
that the particle and hole cluster size distributions in the steady state
decay as a power-law: 
\beq
Prob(l) \sim l^{-{\theta}}
\label{powerlaw}
\eeq
with $\theta \simeq 1.8$.
This allows us to understand some features of the correlation functions
discussed above.  Within the independent interval approximation 
\cite{Satya5}
which ignores correlations in the lengths of successive clusters, the 
two-point
correlation function can be calculated and shown to have the scaling form
of Eq. \ref{splt} \cite{das4}.
The cusp in the scaling function is reproduced as well, with the cusp
exponent  $\alpha$ being related to $\theta$ by $\alpha + \theta =2$
within this approximation. On a qualitative note,
the power law cluster distribution makes the interfaces between phases
broad and structured, so that the argument which leads to the Porod law
does not hold any longer.

The phenomenon of fluctuation-dominated phase ordering, with the 
characteristics
of broad distributions of order parameter fluctuations,
power-law cluster distributions, and cusps in the scaled two-point 
correlation functions, can be shown analytically  to hold for 
a related simple model of surface depth
fluctuations. Both steady state properties and time-dependent coarsening 
correlation functions can be calculated within this model \cite{das4,das34},
and the results bear out the picture of FDPO discussed above.

\section{ The Kinematic Wave Phase}

In the stable phase of the LR model ($r_1 > r_2, ~~~p_2>p_1$),
particle motion and landscape fluctuations work oppositely, so that
the resulting state is disordered. The density of both $\sigma$'s
and $\tau$'s is uniform, but the dynamics of density fluctuations
is interesting, and this is the aspect we focus on.

To begin with, we note that the steady state can be shown
exactly to have a product measure form, provided that
$r_1 = p_2$ and $r_2 = p_1$ (the line marked Pairwise Balance in Fig. 1).
To do this, choose new symbols to denote the values of $\sigma$
and $\tau$: Use $1$ if a site or bond is occupied by a $+$ or $/$
and $0$ for $-$ or $\backslash$. Then the elementary moves reduce to
$100 \rt 001$ and $011 \rt 110$ (with rate $r_1$), and the reverse
moves $001 \rt 100$ and $110 \rt 011$ (with rate $r_2$). Now, in
steady state the total probability flux
into each configuration $C$ is equal to the total flux out of $C$.
A sufficient condition for this is  the condition of pairwise
balance \cite{barma4}: in-fluxes and out-fluxes must
balance in pairs, i.e. for every out-flux  $C \rt C'$, there is a unique
in-flux $C'' \rt C$ such that
$W(C^{''} \brt C) P_{ss}(C^{''}) = W(C \brt C^{'}) P_{ss}(C)$
where $W$'s  are transition rates, and $P_{ss}$'s are steady state
probabilities.  Using intuition gained from particle
hopping models, this condition can be shown to be valid \cite{fourauth} if
$P_{ss}(C) = P_{ss}(C^{''})$.
Thus every allowed configuration is equally likely in steady state,
which implies that $P(C)$ is the product of independent site occupation 
probabilities in the thermodynamic limit. An immediate
consequence is that multi-point correlation functions in the steady state 
decouple into products of one-point functions, allowing us to find the exact
expression for the current of each species:
\begin{eqnarray}
J_1= (r_1 - r_2) \rho_1^o (1 - \rho_1^o) (1 - 2 \rho_2^o)
\label{currA}
\end{eqnarray}
for the $\sigma$-particle current, and a similar expression 
with 1 and 2 interchanged for the $\tau$-particle current.

Hydrodynamic equations for density fluctuations can be derived
using these expressions for the current, on following the
procedure of Section 1.1.  Further, by taking linear combinations of the fields,
one can construct eigenmode  fields $h_{\pm}$ 
which decouple at the linear level. These fields describe wave-like modes
travelling with speeds $c_{\pm}$.
The time evolution of the field $h_{+}$ is governed by the equation
\begin{eqnarray}
{{\partial h_+} \over {\partial t}} &=&
c_+{{\partial h_+} \over {\partial x}}
+ D {{\partial^2 h_+} \over {\partial x^2}}
+ {\rm Quadratic~~ Nonlinearities} 
-{K}({{\partial h_+} \over {\partial x}})
\left[({{\partial h_+} \over {\partial x}})^2
-({{\partial h_-} \over {\partial x}})^2\right] + \eta_+(x,t)
\nonumber \\
\label{h+-}
\end{eqnarray}
and a similar equation holds for $h_-$.
The quadratic nonlinearities in this equation are of the same form as in Eq. 
\ref{coupled}.
The coefficients of these nonlinear terms, as also the coefficient $K$ of
the cubic nonlinearity in Eq. \ref{h+-}, depend on the densities $\rho_1^o$
and $\rho_2^o$.
The new noise terms $\eta_{\pm}$ are also delta-correlated. 
While the fields $h_+$ and $h_-$  are decoupled at the linear
level, they are coupled through the nonlinear terms, so that each mode
can influence the dissipation  of the other.

Some of the coefficients in Eq. \ref{h+-} vanish for certain choices of
densities $\rho_{1}^{o}$ and $\rho_{2}^{o}$. Then,
special symmetries arise in the evolution equations
and the dynamical exponents associated with the
wave modes may change.  Four cases  arise.

(a) $RI$ symmetry: Invariance under up-down reflection ($R$)
$h \rt -h$ and under inversion ($I$) of space $x \rt -x$.

(b) $R{\bar{I}}$ symmetry: Invariance under $h \rt -h$ and {\it
not} under $x \rt -x$.

(c) ${\bar{R}}I$ symmetry: Invariance under $x \rt -x$ and {\it not}
under $h \rt -h$.

(d) ${\bar{R}\bar{I}}$ symmetry: Invariance neither under $x \rt -x$,
nor under $h \rt -h$. 

A term like ${\partial^2 h} \over {\partial x^2}$ obeys $RI$
symmetry. Terms like $({{\partial h} \over {\partial x}})$ and
$({{\partial h} \over {\partial x}})^3$ obey $R{\bar{I}}$
symmetry.  The ${\bar{R}}I$ symmetry is respected by the term
$({{\partial h} \over {\partial x}})^2$, while a term like
$({{\partial h} \over {\partial x}})$ added to it
breaks that and gives rise to ${\bar{R}}{\bar{I}}$ symmetry.

To illustrate the occurrence and effects of different types of symmetries, 
we consider three special pairs of densities ($\rho_1^o$,$\rho_2^o$).

(I) For $\underline{\rho_1^o = \rho_2^o = {1 \over 2}}$, Eq. \ref{h+-}
and its partner for $h_-$ reduce to a pair of coupled equations, 
with linear first and
second derivative terms and cubic gradient nonlinearities. Quadratic
nonlinearities are absent.

The equations describe two kinematic waves moving with speed $c_+=r^{'}/2$ 
and
$c_-=-r^{'}/2$ where $r^{'}=r_1-r_2$. The nonlinear couplings 
imply that each wave influences the
evolution of the other. In order to study the dissipation of say
the $+$mode, it is essential to move to the frame which co-moves with it,
through the Galilean shift $x \rt x + c_+t$, $t \rt t$.

Evidently, in this frame, the $-$mode has a speed $c_{-} - c_{+}$.
The evolution equation is invariant under
$h_+ \rt -h_+$, $h_- \rt -h_-$ but not $x \rt -x$, because of the
linear ${{\partial h_-} \over {\partial x}}$ and cubic nonlinear terms.
The $R{\bar{I}}$ symmetry holds in the rest frame of $h_+$ mode.
Similarly the dissipation of the $-$mode can be
studied by going to a frame which co-moves with it.
It is easily seen that $R{\bar I}$ symmetry holds in this frame as well.

Each of the $+$ and $-$modes thus have the same symmetry as a single scalar
field evolving through Edwards-Wilkinson dynamics with additional cubic 
nonlinearities.  For the corresponding single-field system, 
we have $z = 2$  (though with multiplicative powers of
logarithms) \cite{toom}, and we might expect the same behaviour here.

(II) For \underline{${\rho_1^{o} = 1/2}$ and ${\rho_2^{o} \neq 1/2}$},
on going to either of the frames in which the $+$ mode or the $-$ mode are at
rest, we see that the ${\bar{R}}{\bar{I}}$ symmetry
applies for each of the fields; the $R$ symmetry is broken by
by quadratic nonlinear terms, and $I$ is broken because of linear first
order and cubic terms. The most relevant terms at the linear fixed point
are the quadratic nonlinear terms. Thus we would expect these terms
to govern the dissipation and give rise to the KPZ value $z = 3/2$
\cite{KPZ5} for both the modes.

(III) For $\underline{\rho_1^{o} = \rho_2^{o} \neq 1/2}$,
an interesting situation
arises. In the co-moving frame of the $-$mode, the pair of equations
have symmetry under $h_- \rt -h_-$ and $h_+ \rt h_+$ but not under
$x \rt -x$. Thus the $h_-$ field has $R{\bar{I}}$ symmetry, while the
moving $h_+$ field has ${\bar R}{\bar I}$ symmetry.
The same symmetries hold in the rest frame of the $+$ wave.
Based on these observations, we expect $z=2$ for the $-$mode
(perhaps with multiplicative logarithmic corrections), and $z = 3/2$ for the
$+$ mode.

Thus based on simple considerations of symmetry, we would expect different
types of behaviour in the cases I, II and III.
These expectations are borne out by both numerical simulations and
analytic calculations.
The equal-space unequal-time autocorrelation function $F(t)
\sim t^{\beta}$ for the lattice model was investigated numerically for
the three cases discussed above. Our Monte Carlo results for $\beta$
are in accord with
expectations based on symmetries of the continuum equations. In case I,
we find $\beta=1/4$ with evidence for multiplicative logarithmic corrections 
for both modes. In case II, we find $\beta=1/3$ for both modes, indicating
KPZ behaviour. Finally, in case III, we find that one mode has type I
behaviour, while the other one has type II behaviour, as expected on
grounds of symmetry. 

An analytical treatment of the generalised Langevin equations (\ref{h+-}) 
is straightforward to carry out \cite{fourauth}. The renormalised 
propagators and correlation functions for $h_+$ and $h_-$ are calculated 
in a 1-loop self-consistent scheme. In case III in particular the character 
of the perturbation theory is controlled by the presence of kinematic waves. 
In contrast to what happens in \cite{abjkbsr}, their effect cannot be shifted 
away by Galilean transformations, and we find as expected from the symmetry 
arguments above that $z_+ = 3/2$ and $z_- = 2$. The calculations for the 
other cases (I and II) likewise give results consistent with the 
symmetry arguments above.

\section{ Conclusion}

We conclude by pointing out some connections to other models
and discussing some open questions.

As discussed in Section 2, the strongly phase separated phase
of the LR model is closely related to the corresponding phase of the
permutation-symmetric ABC
model \cite{mukamel24,mukamel34} especially at the symmetric
point of each model where the detailed balance condition is valid.
However, an analog of the threshold case of fluctuation-dominated
phase ordering which occurs in the LR model has not yet been found
in the ABC or related models.  In particular, the permutation
non-symmetric three-component model of Arndt et al \cite{arndt}
shows a transition
from a phase-separated phase to a disordered phase with a large
correlation length \cite{sasamoto}, but the state at the 
transition point does not seem to show FDPO.

The threshold case of the LR model is tantamount to the problem of
hard-core particles sliding down a fluctuating surface. This
leads to the question of whether fluctuation-dominated phase
ordering survives if the surface in question evolves differently.
A study of the problem with KPZ dynamics for the surface shows
that FDPO survives \cite{das34}, though the value of the cusp
exponent changes from that in the EW case.  On the other hand,
with the surface evolving
according to Das Sarma-Tamborenea dynamics \cite{DT}, there is no cusp
in $C(y)$ (Eq. \ref{nonporod}),
and the Porod law is recovered \cite{das34}. Another interesting direction
to investigate is what happens if the falling particles are noninteracting.
Some results in this direction are available from a study of
domain-wall dynamics in a model of growth of binary films \cite{drossel}.

The continuum equations of motion in the wave phase are a one-dimensional 
reduced model of those governing the dynamics of small fluctuations of 
a flux lattice drifting through a clean type II superconductor \cite{aditi}. 
These equations, or more properly the equations implied by (\ref{coupled}) 
for $\partial_x h_1, \, \partial_x h_2$, with some of the couplings set 
to zero, are Burgers-like equations for magnetohydrodynamics (MHD), 
as shown in \cite{abjkbsr}. The dynamics in this kinematic-wave region 
of the phase diagram is largely governed by KPZ-like critical 
exponents (in particular $z = 3/2$). On special loci in the space of 
parameters, however, weak dynamic scaling prevails: if $h_{\pm}$ are the 
eigenmodes of the linearised versions of the equations, then the 
complete nonlinear theory on these loci is governed by $z_+ = 3/2, \, 
z_- = 2$ for $h_+$ and $h_-$ respectively \cite{fourauth}.   
 
Finally, an important open question concerns the behaviour of 
the LR lattice model in higher
dimensions. Recently, it has been shown that a higher-dimensional version
of the ABC model continues to show strong phase separation \cite{mukamel44}.
It would be interesting to investigate analogous questions for all phases of
the LR model as well. So far, the only result along these lines is the
demonstration of product measure along particular loci 
in the disordered phase in higher dimensions \cite{fourauth}.

\section{Acknowledgement} The work summarised in this article
began in collaboration with Rangan Lahiri whose untimely death in 
1998 ended a scientific career of great promise. We dedicate 
this paper to his memory.  


\end{document}